\documentclass[12pt]{article}
\usepackage[cp1251]{inputenc}
\usepackage[russian,english]{babel}
\usepackage{amsfonts,amssymb}
\usepackage{graphicx}
\usepackage{amsmath, amssymb, graphics, setspace}
\usepackage{hyperref}

\begin{document}

\title{The generalized Lemaitre-Tolman-Bondi solutions with nonzero pressure in modeling the cosmological black holes.}
\maketitle
\begin{center}
 E.~Kopteva$^{1,2}$ \quad P.~Jaluvkova$^{1,2}$ \quad Z.~Stuchlik$^2$\\
       $^1${\it Joint Institute for Nuclear Research, Dubna, Russia} \\
       $^2${\it Institute  of  Physics,  Faculty  of  Philosophy  and  Science,
Silesian  University  in  Opava, Czech Republic} \\
\end{center}

\begin{abstract}
In this work, we generalize the results that were obtained to describe the Schwarzschild-like black hole enclosed in the dust cosmological background by use of the mass function method. Developing this approach we find new exact solutions of the Einstein equations, which are the generalization of the Lemaitre-Tolman-Bondi solution for the case of nonzero pressure. On the basis of these solutions we build the models of cosmological black hole at the matter and radiation dominated epoch.   
\end{abstract}

\section{Introduction}
\label{intro}
The question of constructing a consistent model of the cosmological black hole still attracts the interest of cosmologists as far as it is concerned in a wide set of research directions \cite{Giddings}-\cite{Javad}, besides, in the problem of the black hole horizon dynamics \cite{FirouzjaeeMansouri3}, in the problem of interplay between cosmological expansion and local gravity \cite{MoradiFirouzjaeeMansouri},\cite{FaraoniJacques}, in the problem of structure formation in the early universe \cite{Yokoyama} etc.

Under 'cosmological' black hole we mean here the black hole immersed in some cosmological medium which fills the whole universe.

In spite of being entirely criticized, the most mentioned in the literature is the McVittie solution \cite{17} commonly used for description of the cosmological black holes \cite{FaraoniJacques},\cite{18}-\cite{22}. 

The alternative way to describe the black holes within the cosmological surrounding is to use the Lemaitre-Tolman-Bondi (LTB) models for inhomogeneous matter distribution that was fundamentally explored in \cite{Krasinski1}-\cite{Krasinski3}. In the vein of this approach recently the model was obtained for the Schwarzschild-like black hole enclosed in the dust cosmological background by use of the mass function method \cite{KorkinaKopteva},\cite{Ourart1}.

In this work, we generalize the previous result to the case when the black hole possess an electric charge and the cosmological medium along with dust includes an extra component with nonzero pressure.

The paper is organized as follows. First we briefly present the idea of the approach. Then we successively derive the set of exact solutions of the Einstein equations and generalize them. In conclusions section we summarize the results.

\section{The method for obtaining the exact solutions in General Relativity}
\label{sec:1}
It was shown in \cite{Cahill-McVittie} that for the spherically symmetric interval of general form
\begin{equation}\label{metric1}
ds^2=e^{\nu(R,t)}dt^2-e^{\lambda(R,t)}dR^2-r^2(R,t)d\sigma^2
\end{equation}
with $d\sigma^2=d\theta^2+\sin^2\theta d\varphi^2$ being a metric on unit 2-sphere there exists a combination $rR^{\varphi}_{\theta\varphi\theta}$ which is invariant relative to the coordinate transformations concerning $t$ and $R$ coordinates, where $R^{\varphi}_{\theta\varphi\theta}$ is a respective component of the curvature tensor for the metric (\ref{metric1}). Therefore it is possible to introduce the mass-function (or the so-called Misner-Sharp mass) \cite{MisnerSharp}-\cite{Zannias} 
\begin{equation}\label{mf}
m(R,t)=r(R,t)(1+e^{-\nu(R,t)}\dot{r}^2-e^{-\lambda(R,t)}r'^2),
\end{equation}
which will be invariant provided the spherical symmetry is kept.

Further in or discussion we will omit the indication of the dependence $(R,t)$ near the metric coefficient $r$ for brevity, in case it can not cause confusion.

By use of (\ref{mf}) the Einstein equations may be rewritten in much simpler way
\begin{equation}
m'=\varepsilon r^2 r'; \label{ES1} 
\end{equation}
\begin{equation}
\dot{m}=-p_{\parallel} r^2 \dot{r}; \label{ES2}  
\end{equation}
\begin{equation}
2\dot{r}'=\nu ' \dot{r}+\dot{\lambda} r';\label{ES3}    
\end{equation}
\begin{equation}
2\dot{m}'=m'\frac{\dot{r}}{r'}\nu'+\dot{m}\frac{r'}{\dot{r}}\dot{\lambda}-4 r \dot{r} r' p_{\perp} .  \label{ES4}
\end{equation}
Here and further we use the system of units where velocity of light $c=1$ and factor $8 \pi G=1$; dot and prime mean partial derivatives with respect to $t$ and $R$, respectively; $\varepsilon$ is energy density, $p_{\perp}$ is tangential pressure, $p_{\parallel}$ is radial pressure. 

The metric describing the inhomogeneous matter distribution in comoving coordinates written in rather general form reads
\begin{equation} \label{comovmetric} 
ds^{2} =dt^{2} -\frac{r'^{2} (R,t)}{f^{2} (R)} dR^{2} -r^{2} (R,t)d\sigma ^{2} , 
\end{equation}
where $f(R)$ is an arbitrary function. In this paper we will consider the space of zero curvature that yields  $f(R)=1$.
The mass function (\ref{mf}) in this case will take the form
\begin{equation} \label{flatmf} 
m(R,t)=r(R,t)\dot{r}^{2}(R,t). 
\end{equation} 

Regarding the equation (\ref{ES1}) the mass function may be tentatively interpreted as the total mass (including gravitational energy) enclosed in between the center of the distribution and the 2-space of symmetry passing through the point $\lbrace t,R,\theta,\varphi \rbrace$. Due to that interpretation in some cases the mass function posses the property of additivity. So in case of negligible interaction one can obtain the solution of Einstein equations (\ref{ES1})-(\ref{ES4}) for the system that may contain several sources. 

By use of this idea the solution describing the Schwarzschild-like black hole in the dust cosmological medium was obtained in \cite{KorkinaKopteva} as the LTB solution with combined mass function. Further we will generalize this approach for some different situations such as the Reissner-Nordstrem black hole embedded into the dust background, neutral black hole in the medium that contains both dust and radiation, and finally the Reissner-Nordstrem black hole immersed in the universe filled with dust and radiation.  

\section{The solution describing the black hole in the universe filled with dust and radiation}
\label{sec:2}
The exact homogeneous solution for the universe filled simultaneously with dust and radiation was obtained by Chernin in \cite{Chernin}. For the comoving interval (\ref{comovmetric}) under zero curvature the Chernin solution has the form:
\begin{equation} \label{Cherninsol1} 
r=R\left(\frac{a_{dust}}{4}\eta^2+a_{rad}\eta\right),
\end{equation}
\begin{equation} \label{Cherninsol2} 
t-t_0=\frac{a_{dust}}{12}\eta^3+\frac{a_{rad}}{2}\eta^2,
\end{equation} 
where $t_0$ is an arbitrary constant and $\eta$ is a parameter. The mass function for this case according to the definition (\ref{flatmf}) will be
\begin{equation} \label{mfdustrad} 
m_{dr}(R,t)=a_{dust} R^3 + \frac{a_{rad}^2 R^4}{r}, 
\end{equation}
where $a_{dust}$ and  $a_{rad}$ are constants related to the size of the universe in particular cases of pure dust and pure radiation, respectively. The first term in (\ref{mfdustrad}) is exactly the mass function in the flat Friedman solution for the dust, the second term is the mass function in the Friedman-like solution for the radiation. 
It can be easily shown by direct calculation that the mass function for the Schwarzschild solution is $m_S=r_g$, where $r_g$ is the Schwarzschild radius.

Let us now consider the black hole with the mass function $m_S$ immersed in the cosmological background represented by the mixture of non-interacting radiation and dust matter which is described be the mass function (\ref{mfdustrad}). Supposing the weak interaction between the black hole and the cosmological background we will combine the mass function for the system in the following way:
\begin{equation} \label{mfdrs} 
m(R,t)=r_g+a_{dust} R^3 + \frac{a_{rad}^2 R^4}{r}. 
\end{equation}

The comoving interval in this case will be still defined by the expression (\ref{comovmetric}), while the spatial coordinate $R$ and metric coefficient $r(R,t)$ there may somehow change. As far as the mass function is invariant relative to $\lbrace R,t \rbrace $ transformations it will keep its structure, and using the definition (\ref{flatmf}) one can obtain the following equation for the metric coefficient $r(R,t)$:          
\begin{equation} \label{eqndrs} 
\dot{r}=\pm\frac{\sqrt{r}}{\sqrt{r_g+a_{dust} R^3 + \frac{a_{rad}^2 R^4}{r}}}, 
\end{equation}
which may be easily integrated leading to the result:
\begin{equation} \label{soldrs} 
t - t_0(R) =\pm \frac{2}{3} r^{3/2} \frac{\sqrt{m}\left(m - \frac{3 a_{rad}^2 R^4}{r} \right)}{\left(r_g+a_{dust} R^3 \right)^2}, 
\end{equation}
where $t_0(R)$ is an arbitrary function of integration and $m$ is the mass function given by (\ref{mfdrs}); $\pm$ corresponds to the case of the universe expansion/contraction, respectively.

In particular case, when $a_{rad}=0$ the solution (\ref{soldrs}) turns to the solution for the black hole enclosed in pure dust that was obtained in \cite{KorkinaKopteva} and investigated in \cite{Ourart1}:
\begin{equation} \label{sold} 
r=\left[\pm\frac{3}{2} \sqrt{r_{g} + a_{dust} R^{3} } (t-t_{0} (R))\right]^{2/3}. 
\end{equation}  

The solution (\ref{soldrs}) as well as the solution (\ref{sold}) contains the Schwarzschild black hole in the central region ($ R<<1 $). It can be seen by expanding of (\ref{soldrs}) into a series near $R \sim 0$. Keeping the accuracy up to the second order one obtains:
\begin{equation} \label{limitsoldrs} 
r \approx \left[\pm\frac{3}{2} \sqrt{r_g}( t - \tilde{t_0}(R))\right]^{2/3}. 
\end{equation}
This is exactly the flat Schwarzschild solution in synchronous coordinates with certain choice of the arbitrary function $\tilde{t_0}(R)=t_0(0)+ t'_0(0) R +\frac{1}{2} t''_0(0) R^2$.

\section{The solution describing the charged black hole immersed in the dust background}
\label{sec:3} 
Let us now consider the situation when the electrically charged black hole is immersed in the universe filled with the dust matter.

The space-time around the black hole with nonzero electric charge $q \neq 0$ in the empty space is described by the known Reissner-Nordstrom solution
\begin{equation} \label{RN} 
ds^{2} =\left(1-\frac{r_{g} }{r} + \frac{q^2}{r^2}   \right)dt^{2} -\left(1-\frac{r_{g} }{r} +\frac{q^2}{r^2}  \right)^{-1} dr^{2} -r^{2} d\sigma ^{2} . 
\end{equation} 
According to the definition (\ref{mf}) the mass function for this solution will be
\begin{equation} \label{mfRN} 
m_{RN}(r)= r_g - \frac{q^2}{r}. 
\end{equation}

The background universe in this case is the flat Friedmann universe with the familiar mass function $m_d(R)=a_{dust} R^3$.

Regarding again the comoving metric (\ref{comovmetric}) with $f(R)=0$ we will combine the mass function taking into account its invariance relative to $\lbrace R,t \rbrace $ coordinate transformations. The resulting mass function for the studied system will be
\begin{equation} \label{mfRNd} 
m(R,t)= r_g + a_{dust} R^3 - \frac{q^2}{r}. 
\end{equation}   
Applying the relation (\ref{flatmf}) we derive the following equation for the metric coefficient $r(R,t)$
\begin{equation} \label{eqnRNd} 
\dot{r}=\pm\frac{\sqrt{r}}{\sqrt{r_g+a_{dust} R^3 - \frac{q^2}{r}}}, 
\end{equation}
that can be easily integrated:
\begin{equation} \label{solRNd} 
t - t_0(R) =\pm \frac{2}{3} r^{3/2} \frac{\sqrt{m}\left(m + \frac{3 q^2}{r} \right)}{\left(r_g+a_{dust} R^3 \right)^2}, 
\end{equation}
with $m$ given by (\ref{mfRNd}).

Again in the limit of $q=0$ the solution (\ref{solRNd}) turns into the mentioned solution (\ref{sold}).

Taking the limit of small $R$ one will obtain
\begin{equation} \label{limitsolRNd}
t - \tilde{t_0}(R) =\pm \frac{2}{3} \frac{\sqrt{r_g r- q^2}\left(r_g r  + 2 q^2 \right)}{r_g^2}. 
\end{equation}   
This expression exactly represents the flat case of the Reissner-Nordstrom solution in synchronous coordinates with certain choice of the arbitrary function $\tilde{t_0}(R)=t_0(0)+ t'_0(0) R +\frac{1}{2} t''_0(0) R^2$.

\section{The generalized case for the charged black hole immersed in the universe filled with dust and radiation}
\label{sec:4} 
Let us now consider the electrically charged black hole embedded into the cosmological background which includes the dust matter and radiation simultaneously. The metric describing the resulting space-time in comoving coordinates for the flat space case will be still the metric (\ref{comovmetric}). The combined mass function of the whole system will be
\begin{equation} \label{mfRNdR} 
m(R,t)= r_g + a_{dust} R^3 + \frac{a_{rad}^2 R^4}{r} - \frac{q^2}{r}. 
\end{equation}
Similarly to the previous cases involving the definition (\ref{flatmf}) we obtain the equation for the metric coefficient $r(R,t)$
\begin{equation} \label{eqnRNdR} 
\dot{r}=\pm\frac{\sqrt{r}}{\sqrt{r_g+a_{dust} R^3 + \frac{a_{rad}^2 R^4}{r} - \frac{q^2}{r}}}, 
\end{equation}   
and its integral
\begin{equation} \label{solRNdR} 
t - t_0(R) =\pm \frac{2}{3} r^{3/2} \frac{\sqrt{m}\left(m + 3 \frac{q^2 - a_{rad}^2 R^4}{r} \right)}{\left(r_g+a_{dust} R^3 \right)^2}, 
\end{equation} 
where $m$ is the total mass function (\ref{mfRNdR}). The reasoning as in previous sections about the limits of the solution (\ref{solRNdR}) remains to be true. The central part $(R<<1)$ of the solution is either the Reissner-Nordstrem black hole or the Schwarzschild one provided electric charge is zero.
All the solutions (\ref{soldrs}), (\ref{sold}), (\ref{solRNd}) discussed above are the particular cases of (\ref{solRNdR}).

Some notions about avoiding the shell-crossing singularities in solutions of such kind can be found in \cite{Ourart1}. In this paper everywhere we supposed the right choice of the arbitrary function $t_0(R)$ so that all its derivatives behave well at $R=0$.

The advantage of the solution (\ref{solRNdR}) is obvious, as far as it has very clear physical interpretation and it is the exact solution which naturally includes the respective limit cases. The problem of the black hole horizon in the mentioned systems may be explored on the basis of this solution if one makes the transition to the curvature coordinates. The general consideration of the horizons behavior and singularities in the LTB solutions can be seen in \cite{Krasinski2}.

\section{Conclusions}
\label{concl}
In this work we have obtained the set of new exact solutions (\ref{soldrs}), (\ref{solRNd}), (\ref{solRNdR}) of the Einstein equations that generalize the known LTB solution for the particular case of nonzero pressure under zero spatial curvature. These solutions are pretending to describe the black hole immersed in nonstatic cosmological background and give a possibility to investigate the hot problems concerning the effects of the cosmological expansion in gravitationally bounded systems. The solution (\ref{soldrs}) may be used as a seed model in the problem of structure formation in the universe at the epoch of matter and radiation decoupling.

It was shown that each of the solutions obtained contains either the Reissner-Nordstrom or the Schwarzschild black hole in the central region of the space.

It is demonstrated that the approach of the mass function use allows clear physical interpretation of the resulting solutions, that benefits much any their concrete application.

\end{document}